\pdfoutput=1

\documentclass[letterpaper, 10pt, conference]{ieeeconf}  

\usepackage[compact]{titlesec}
\usepackage{amssymb,amsmath,amsthm}
\usepackage[compact]{titlesec}
\usepackage{cite}
\usepackage{amsmath,amssymb,amsfonts}
\usepackage{algorithmic}
\usepackage{graphicx}
\usepackage{textcomp}
\usepackage{xcolor}
\usepackage{amsmath}
\usepackage{multirow} 
\let\labelindent\relax
\usepackage[shortlabels]{enumitem} 
\IEEEoverridecommandlockouts                              

\title{\LARGE \bf
Event-Driven Real-Time Multi-Objective Charging Schedule Optimization For Electric Vehicle Fleets
}

\author{Jose Peeterson Emerson Raja$^{1}$ and Arvind Easwaran$^{2}$
\thanks{“This study is supported under the RIE2020 Industry Alignment Fund – Industry Collaboration Projects (IAF-ICP) Funding Initiative, as well as cash and in-kind contribution from the industry partner(s).”}
\thanks{ Authors belong to Continental-NTU Corporate Lab,  Nanyang Technological University, 50 Nanyang Avenue, 639798, Singapore.
       Emails: {\tt\small josepeeterson.er@ntu.edu.sg} and {\tt\small arvinde@ntu.edu.sg}}\
}

\begin{document}
\setlength{\abovedisplayskip}{5pt}
\setlength{\belowdisplayskip}{5pt}

\maketitle
\thispagestyle{empty}
\pagestyle{empty}

\begingroup\renewcommand\thefootnote{\textcopyright}
\footnotetext{202X IEEE. Personal use of this material is permitted. Permission from IEEE must be obtained for all other uses, in any current or future media, including reprinting/republishing this material for advertising or promotional purposes, creating new collective works, for resale or redistribution to servers or lists, or reuse of any copyrighted component of this work in other works.}
\endgroup

\begin{abstract}

The utilization of Electric Vehicles (EVs) in car rental services is gaining momentum around the world and most commercial fleets are expected to fully adopt EVs by 2030. At the moment, the baseline solution that most fleet operators use is a Business as Usual (BAU) policy of charging at the maximum power at all times when charging EVs. Unlike petrol prices that are fairly constant, electricity prices are more volatile and can vary vastly within several minutes depending on electricity supply which is influenced by intermittent energy supplies like renewable energy and increased demand due to electrification in many industrial sectors including transportation. The battery in EVs is the most critical component as it is the most expensive component to replace and the most dangerous component with fire risks. For safe operation and battery longevity it is imperative to prevent battery capacity fade whenever the EVs are under the control of the fleet operator such as during charging. Fundamentally, the fleet operator would like to service as much demand as possible to maximize the revenue generated at a particular time instance. This is achieved by minimizing the EV's time spent on charging and thereby increasing their availability for rides. The three goals of reducing charging cost, battery capacity fade and maximizing ride availability are formulated as a multi-objective optimization problem. The formulation is tested using the Gurobi solver on two cases from the real-world ACN dataset involving low and high EV charging densities over a week long period. The results of the proposed solution show 33.3\% reduction in peak electricity loading period, 53.2\% savings in charging cost and 16\% lower battery capacity fade for the fleet operator.

\end{abstract}

\section{INTRODUCTION}
\subsection{Background}
Many governments around the world have a target to phase out Internal Combustion Engine vehicles (ICE) completely. The Singapore Green Plan 2030 \cite{b1} envisions up to 60,000 EV charging points by 2030. By the year 2040, the Land Transport Authority of Singapore wants to reduce carbon emissions by 1.5 to 2 million tonnes \cite{b2}. Consequently, this will lead to an influx of EVs which will add more strain to the electricity supply network. Peak demand and intermittent energy supply are expected to cause highly fluctuating electricity prices which makes uncontrolled charging cost-ineffective.

Additionally, special care must be taken for the safe operation of Lithium-ion batteries in the EVs. The battery capacity experiences different levels of fade for different operating conditions. Lower battery capacity translates to a shorter driving range which exacerbates the range anxiety problem. Individual EV capacity losses also add up to cause significant fleet level capacity losses which results in more frequent charging of the EVs, less time spent servicing rides and a monetary loss in the value of the EVs.

During peak customer demand when the profits from servicing rides far outweigh the benefits of saving on charging cost and battery capacity fading cost in the near term, a quick charging policy that does not violate the charging infrastructure limits and charges EVs in the order of earliest to latest departure times is preferred.

An EV rental fleet operator usually has hundreds of EVs serving a city with Charging Stations (CS) distributed across it. A customer reserves a vehicle, $v$ in advance for use at a future departure time, $ T^{dep}_v $ for a journey requiring a certain amount of departure State of Charge (SoC), $ SoC^{dep}_{v} $. The plugged vehicles already at the CS have an initial SoC, $ SoC_{-1,v} $ at the beginning of optimization implementation. A newly arrived EV at time, $ T^{arr}_v $ has an arrival SoC,  $SoC^{arr}_{v} $. After picking-up, the vehicles can be used till the current SoC, $ SoC_{v}^{cur}$ reaches a lower threshold such as 40\%. Upon returning the vehicle, it is plugged back to recharge.

Typically, most fleet operators recharge their vehicles at maximum power hoping to ensure that they meet all possible future ride requests. This is used as the baseline solution and it does not do a cost-benefit analysis of the near term.

The solution to this problem is then to generate a charging schedule in real-time that is robust to dynamic EV arrival and departure events at a CS. The schedule should also ensure that all the EVs receive the departure state of charge before their departure time while simultaneously optimizing the three objectives of charging cost minimization, battery capacity fade minimization and ride availability maximization in order to benefit the fleet operator and reduce peak loading period of the grid.

\subsection{Related work}

In this section existing literature on the impact of EV charging on the grid and the three main concerns of the fleet operator including charging cost minimization, battery capacity fade minimization and ride availability maximization are reviewed.

The negative impact of uncontrolled and uncoordinated charging of EVs on the grid has been analyzed by \cite{b3}. These include voltage and frequency variations, higher copper losses and overloading of transformers and transmission lines. Thus, coordination of charging operation between the EVs is advantageous for normal grid operation. In \cite{b4}, a model predictive control (MPC) based method was used by the aggregator to determine the real-time decisions of the EV charging and volumes for the aggregator to transact in the regulating power market and fulfilling the charging requirements of the EVs. However, the charge scheduling of the EVs was made a day-ahead so it did not give enough flexibility in view of dynamic ride requests faced by fleet operators. 

\subsubsection{Electricity Charging Cost}

The electricity charging cost is a major operational cost of any EV fleet operator. The charge in EV's batteries deplete even in the idle state. Continuously charging EVs leads to energy losses. Saving electricity cost is made possible by the fact that many suppliers such as the Energy Market Company in Singapore provide forecasted electricity price two days into the future \cite{b5}. In \cite{b6} the monetary cost of a parking lot that charged EVs using both conventional and photovoltaic generation was minimized through a mixed integer linear programming problem to obtain the charging schedule of the EVs. Most fleet operators however, prefer to use electricity provided by mainstream suppliers as standalone photovoltaic generation is small, unstable and needs to be self-managed. In \cite{b7}, the economic benefits to EVs, CSs and retailers was forumalated as an equilibrium problem. A distribution system operator collected information from all the participants and performed day-ahead scheduling and then conveyed the results back to act upon. Day-ahead forecasted electricity prices are less accurate than prices forecasted every few hours due to frequent price revisions based on instantaneous demand and supply. To mitigate this problem they used fast chargers to minimize the time during charging and reduce the impact of price variations on the next day. 

\subsubsection{Battery Capacity Fading Cost}

Lithium iron phosphate (LiFePO4) battery capacity fading has been studied in \cite{b8} - \cite{b10} to identify stress factors such as charging current, temperature, level of charge and the depth of discharge that caused cyclic and calendric capacity fade during uncontrolled charging events. These studies showed that the battery should only be charged enough for the upcoming trip together with an additional charge of approximately 10\% to prevent over-discharge. 
In \cite{b11} Electric bus fleets (EBF) were scheduled through dynamic programming to minimize battery replacement costs over the service life of the EBFs. They considered the battery capacity fading process and working loads of the buses for different routes and used a reverse order matching strategy to assign buses to their corresponding routes. Commercial EV fleets however, have similar working loads in an urban environment. A simplistic capacity fading model only based on the Arrhenius relationship was used to measure the capacity fading process. 
In \cite{b12} two control strategies called Variable Initial Charging Time Scheme and Adjustment of Charging Power Strategy where the EV charging process waited and charged by a constant power during high loads and high power during low loads. This was found to prolong battery life and smoothen grid load fluctuations. However, this is considered to be a static policy that does not consider the detailed effects of stress factors of a capacity fading model.
The effect of charging strategies including uncontrolled and demand side management (DSM) on battery capacity fading was analyzed for private EVs on a danish island in \cite{b13}, They showed that smart charging strategies reduced the annual average SoC to 12–17\% due to charging being postponed closer to departure time. In this application, the EVs were mainly parked most of the day and calendric fading contributed more than cyclic fading. Cyclic capacity fading is more common in commercial EVs. In \cite{b18}, a battery capacity fading cost model for real-time analysis was incorporated into optimizing economic power dispatch in a system involving renewable energy sources, charging coordination of EVs and vehicle to grid. In the battery model which only considered cyclic capacity fading, the less accurate stress factor called the number of charge-discharge cycles was used instead of the charge processed in a cycle. In \cite{b20}, the charging power in a park-and-charge business was regulated in order to minimize the battery capacity fading cost of customers and the operating cost of the business. The battery capacity fading model was purely based on temperature and no other stress factors. charging power was only related to the temperature though a simple linear relationship involving the thermal resistance of the battery pack. These reviewed papers show that there is a lack of detailed modelling of the battery capacity fading process in the multi-objective optimization setting.

\subsubsection{Ride Availability Utility}

As payment from rides is the main source of income for fleet operators, reducing charging time and thus increasing the available time for rides is preferable. In \cite{b14} a scheduling management system was used to avoid waiting time by informing the driver of the number of EVs waiting at CSs. This passive approach was used to prevent long waiting times at CS by initially recommending stations to customers to charge at stations with fewer charging vehicles thereby distributing the quantity of EVs more evenly amongst CSs. And also reducing the quantity of waiting vehicles and preventing clustering at a single station. In this strategy, the charging time of individual EVs was not directly controlled by adjusting the power to the EVs in the order of their deadlines. Performance from this indirect approach was inconsistent as it depended on driver preferences.
In \cite{b15} a server-based scheduling management system recommended charging station to EVs in order to optimize the economic cost and reduce charging time by minimizing queues at charging stations using an Arrival time-based priority (ATP) algorithm. This algorithm allotted the charging point based on the arrival time and updated the waiting times for other EVs. The departure times of the next trip were not accounted and there was no notion of priority. The EVs heeded to recommendations only due to penalties imposed for acting otherwise. The problem of EVs taking too long to charge was tackled in \cite{b16} by adding an occupation fee term to the charging cost objective. However, the occupation fee was chosen arbitrarily high or low without considering the exact price and ride availability. In \cite{b17}, an agent based transport model generated the number and position of charging stations needed to serve customers for a fixed waiting time. It was found that to meet the waiting time and be profitable double the number of short range vehicles that were charged with slow charging were required compared to long range vehicles that could be charged by fast charging. Clearly, this model led to the fleet operator having to procure and manage more EVs.

In order to optimize the aforementioned conflicting objectives where optimizing one causes deterioration of another, the multi-objective optimization (MOO) technique is employed. In \cite{b19} minimization of electricity cost, battery degradation, grid net exchange and CO2 emissions were performed. In \cite{b24} the conflicting objectives of EVs, CSs and electricity suppliers were optimized simultaneously by modelling and solving them as an equilibrium problem iteratively. The solution resulted in the day-ahead charging schedule. In both cases the optimization was not dynamic to real-time events. 

\subsection{Main contributions}

To address these dynamic concerns a decentralized online scheduling strategy at the EV's CS level is implemented. A multi-objective optimization framework at a particular CS performs these optimizations. The main contributions are:

1) Simultaneously optimizing the charging cost, battery capacity fade and ride availability in order to maximise the profits for the fleet operator and reduce grid congestion over the near future. The online optimization is robust to dynamic EV arrivals and departures and the charging schedule is generated in real-time.

2) Complete modelling of cyclic and calendric battery capacity fading using empirical data. Simplification of a non-linear capacity fading model in order to solve optimization problems rapidly and obtain real-time performance.

3) Verification and validation of improved performance of our proposed solution compared with the baseline on two cases of a real-world charging pattern of EVs by implementing and solving using Gurobi.

This paper is organized as follows. In Section II the battery capacity fading is rigorously modelled. This is followed by the mathematical formulation of the multiple objectives together will all the constraints defined in section III. Finally, Section IV presents details of the simulation and analysis of results followed by conclusions drawn in section V.

\section{Battery capacity fading model}
The cell chemistry with graphite anode and LiFePO4 cathode is considered in modelling battery capacity fading as it is the most widely used battery for EVs. The cyclic capacity fading  model is obtained from \cite{b8}.

\subsection{Cyclic capacity fade}

The cyclic capacity fading model is shown in (1):
\begin{align}
    C_{loss,cyc,i,v} = [ ( k_1 \cdot SoC_{dev,i,v} \cdot e^{k_2 \cdot SoC_{avg,i,v}} \nonumber \\
    + k_3 \cdot e^{k_4 \cdot SoC_{dev,i,v}}  ) \cdot e^{ \frac{-E_a}{R} \cdot \frac{1}{T_{i,v} - T_{amb}} } ] \cdot {Ah}_{i,v} \ \ \ \ \ \ \ \ \ \ \ 
\end{align} \\
\\
The stress factors during charging are the average state of charge, $SoC_{avg,i,v}$, the depth of discharge from the average SoC, $SoC_{dev,i,v} $ and the charge processed, $Ah_{i,v}$ for a vehicle $v$ in time slot $i$. $C_{bat}$ is the battery's nominal capacity. All the three stress factors are a linear function of the charging current, $I_{i,v} $. A constant current which results in the charge processed, $Ah$ is allocated in each time slot for a vehicle results in the state of charge, $SoC$ increasing linearly from its initial SoC, $ SoC_{init}$. $SoC_{avg,i,v}$ and $SoC_{dev,i,v}$ are computed as follows.\\

$ SoC = SoC_{init} + \frac{Ah}{C_{bat}} $ \\

$ SoC_{avg} = \frac{1}{Ah} \cdot \int_{0}^{Ah} SoC_{init} + \frac{Ah}{C_{bat}} \,d{Ah}   $ 

\begin{equation}
SoC_{avg} = SoC_{init} + 0.5 \cdot \frac{Ah}{C_{bat}}
\end{equation}

$ SoC_{dev} = \sqrt{ \frac{3}{Ah} \cdot \int_{0}^{Ah} \left( SoC(Ah) - SoC_{avg} \right)^2 \,d{Ah} } $ \\

$ SoC_{dev} = \sqrt{ \frac{3}{Ah} \cdot \int_{0}^{Ah} \left( SoC_{-1} + 0.5 \cdot \frac{Ah}{C_{bat}} \right)^2 \,d{Ah} } $ \\

$ = \sqrt{ \frac{3}{Ah} \cdot \left[ \frac{Ah^3}{12\cdot C_{bat}^{2}} \right]^{Ah}_0 }$

$ = \sqrt{  \left[ \frac{Ah^2}{4\cdot C_{bat}^{2}} \right]^{Ah}_0 }$
\parindent0pt
\begin{equation}
SoC_{dev} =  0.5 \cdot \frac{Ah}{C_{bat}}
\end{equation}

$E_a$ and $R$ are the activation energy and gas constant. $T_{i,v}$ and $T_{amb}$ are the battery and ambient temperatures. The battery's cyclic capacity fading model coefficients $k_1$, $k_2$, $k_3$ and $k_4$ was obtained by fitting (1) to the Stanford battery degradation dataset \cite{b22} using the Matlab surface fitting tool. The cyclic capacity fade prediction was tested on cells and it had a root mean squared percentage error of 3.7\%.

The non-linear cyclic capacity fading model, $C_{loss,cyc,i,v}$ as a function of the charging current $ I_{i,v}$ and the initial SoC, $SoC_{init}$ is given in (4). $SoC_{init}$ is the initial SoC at each time slot. $\Delta t $ is the length of a time slot.
\begin{align}
C_{loss,cyc,i,v} = \left[\left( k_1 \cdot \frac{0.5 \cdot I_{i,v} \cdot \Delta t}{C_{\text{bat}}} \cdot e^{k_2 \left(SoC_{init} + \frac{0.5 \cdot I_{i,v} \cdot \Delta t } {C_{\text{bat}}}\right)} \right. \right. \nonumber \\
\left. \left. + k_3 \cdot e^{k_4 \cdot \frac{0.5 \cdot I_{i,v} \cdot \Delta t}{C_{\text{bat}}}} \right)\cdot e^{-\frac{E_a}{R}\left(\frac{1}{T_{i,v}}-\frac{1}{T_{amb}}\right)} \right] 
 \cdot ( I_{i,v} \cdot \Delta t )^{0.5} \ \ \ \ \ \ \ 
\end{align} 
\\

From visualizing the plot in Fig. 1, two disjoint pieces of the surface plot can be distinguished by splitting the domain on either side of the line  $ I_{i,v} = 480 \cdot SoC_{init} $. A second order polynomial function in (5) with two different set of coefficients given in table 1 are used to approximate the two regions. The goodness of fit parameter, R-square on the domain $I < 480*SoC_{init}$ was 0.9992 and 0.9983 on the domain  $I \geq 480*SoC_{init}$. This approximation is necessary to ensure real-time performance in optimization calculations as the branch and bound algorithm in Gurobi takes too long to converge with the original non-linear model.

\begin{figure}[htbp]
\centerline{\includegraphics[width=8cm,height=6cm]{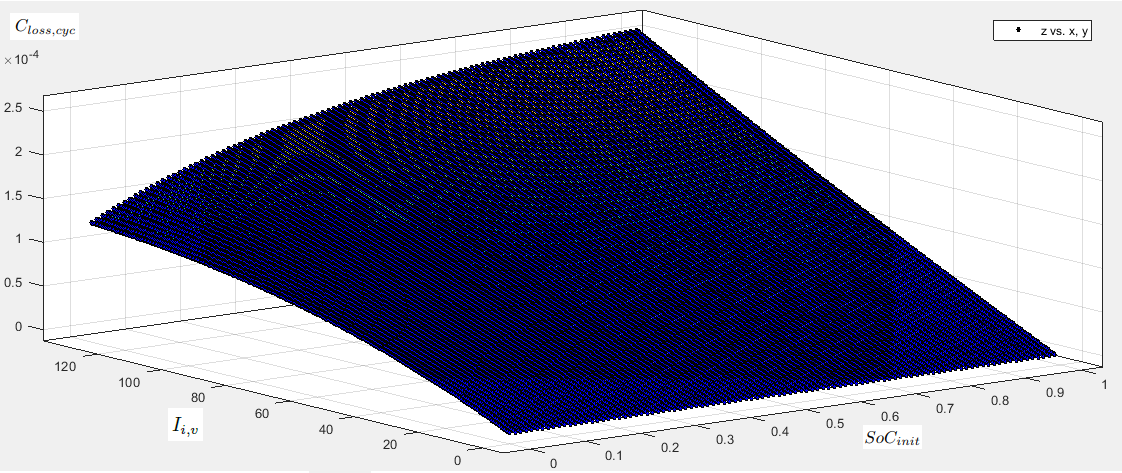}}
\caption{Surface plot of $C_{loss,cyc}$ vs. $SoC_{init}$ vs. $ I_{iv} $}
\label{fig}
\end{figure}

\begin{equation}
\begin{aligned}
C_{loss,cyc,i,v} \approx p_{00} + p_{10}\cdot SoC_{avg,i,v} + p_{01}\cdot I_{i,v} \\
	+ p_{11} \cdot SoC_{avg,i,v}\cdot I_{i,v} + p_{02}\cdot I_{i,v}^2 
\end{aligned}
\end{equation} \parindent0pt

\begin{table}[htbp]
\caption{Coefficient values on different domains}
\begin{center}
\begin{tabular}{|c|c|c|}
\hline
\textbf{}&{\textbf{ $ I \geq 480 \cdot SoC_{init} $}}&{\textbf{ $ I < 480 \cdot SoC_{init} $}}\\
\hline
\textbf{Model Coefficient} & \textbf{\textit{Coefficient value }}& \textbf{\textit{Coefficient value }} \\
\hline
p00 &  4.169*10**-6  $^{\mathrm{a}}$& 6.886*10**-6 \\
\hline
p10 & -9.871*10**-5 $^{\mathrm{a}}$& -1.075*10**-5  \\
\hline
p01 &  1.63*10**-6 $^{\mathrm{a}}$& 1.361*10**-6 \\
\hline
p11 & 2.661*10**-6 $^{\mathrm{a}}$& 6.348*10**-7 \\
\hline
p02 &  -5.757*10**-9 $^{\mathrm{a}}$& -1.902*10**-10 \\
\hline
\end{tabular}
\label{tab1}
\end{center}
\vspace{-6mm}
\end{table}

\subsection{Calendric capacity fade}
The calendric capacity fading model was obtained from \cite{b21} and is shown in (6).
\parindent0pt
\parindent0pt
\begin{equation}
\begin{aligned}
 C_{loss,cal,i,v} = C_{bat} \cdot (( \frac{K(T_{amb},SoC_{i,v}) \cdot (1 + \alpha(T) ) \cdot t  }{C_{bat}} \\ + 1 )^{\frac{1}{1+\alpha(T)}} - 1 )
\end{aligned}
\end{equation}
\parindent0pt

In (6) $K(T_{amb},SoC_{i,v})$ and $(1 + \alpha(T)) $ are the kinetic dependence and temperature dependent alpha parameter. $t$ is the time for which the operating conditions are maintained.  Calendric capacity fade, $C_{loss,cal}$ versus $SoC$ has a nearly linear relationship so it is approximated by (7). Coefficients for $p_1$ and $p_2$ are 0.0001347 and 0.00005356 respectively.
\hfill 
\parindent0pt
\begin{equation}
\begin{aligned}
 C_{loss,cal,i,v} \approx p_1\cdot SoC_{avg,v,i} + p_2 \end{aligned}
\end{equation}

\subsection{Total capacity fade}
The exact total capacity fading model is the sum of cyclic and calendric capacity fading models given in (4) and (6) while the approximate total capacity fading model used in the simulation is the sum of (5) and (7).

\section{Mathematical formulation}

\subsection{Weighted sum of multi-objectives:  }

\parindent0pt
\begin{equation}
\begin{aligned}
	\min_{I_{i,v}} \ \sum^{TT_v}_{i}\sum^{N_v}_{v} \alpha_1 \cdot ( (WEP_{i})\cdot I_{i,v}
	\cdot V \cdot \Delta t ) + \\ \alpha_2 \cdot (\ C_{\text{loss, tot,i,v}})
	+  \alpha_3 \cdot (- w_{i,v} \cdot I_{i,v} \cdot V)             
\end{aligned} \ 
\end{equation}

The weight $\alpha_1$ in (8) multiplies the charging cost objective which is the sum of products of forecasted wholesale electricity price, $WEP_{i}$ with energy supplied during the charging period. The charging period is the elapsed number of time slots, $TT_v$ between the starting time, $ t_s $ of the simulation and the departure time, $ T^{dep}_v $. With this information, lower current will be consumed during high electricity prices and vice versa. The total charging cost of all the vehicles over their charging periods is considered as an objective function that has to be minimized. The constraints pertaining to all objectives is given in (9) - (13).

\begin{equation}  0  \leq  I_{i,v} \leq I_{max}  \ \ \forall i,v\\ \end{equation}
\parindent0pt
\begin{equation} 0 \leq \sum_{v} {I_{i,v}} \leq I_{c,max} \ \ \ \forall i \end{equation}
\parindent0pt
\begin{equation}  \lceil \frac{T^{dep}_{v} - t_s}{ \Delta t} \rceil = TT_v  \ \ \ \forall  v
\end{equation}
\parindent0pt
\begin{equation} \sum_{i}^{TT_{v}} {I_{i,v} \cdot \Delta t} \geq (SoC^{dep}_{v} - SoC_{-1,v}) \cdot C_{bat} \ \ \ \forall v\end{equation}
\parindent0pt 
\begin{equation} \sum_{i}^{TT_{v}} {I_{i,v} \cdot \Delta t} \leq (SoC^{dep}_{v} - SoC_{-1,v}) \cdot C_{bat} + SoC_{xtra} \ \ \ \forall v \end{equation}
\parindent0pt

$WEP_i$ in (16) is the dollar cost per kilowatt hour (KWh) given in \cite{b23}. The constraints in (9) and (10) are current limits in each individual time slot and CS current limits respectively. (11) calculates the number of time slots in a charging period. (12) and (13) ensure that departure SoC is reached within the deadline.

The weight $\alpha_2$ multiplies the total battery capacity fading objective, $C_{loss, tot,i,v}$ which is the sum of approximate $C_{loss,cyc,i,v}$ and $C_{loss,cal,i,v}$ given by (5) and (7). Battery capacity fade is minimized by considering the stress factors in the battery capacity fading models \cite{b8} and \cite{b21}. In general, high SoC levels, $SoC_{avg,i,v} $ high depth of discharge, $SoC_{dev,i,v} $ and high charging currents, $ I_{i,v} $ lead to higher fading. The battery should only be charged with enough energy for the upcoming trip. Therefore, charging the battery gradually later on near the end of the charging period reduces the SoC levels, depth of discharge and charging currents throughout the charging period. Minimizing the calendric and cyclic capacity fading models achieves this objective.  The capacity losses are linked to the charging current through the state of charge and charge processed equality constraints in (14) - (17).
\parindent0pt
\begin{equation}
  SoC_{i,v} = SoC_{i-1,v} + \frac{I_{i,v}  \ \cdot \Delta t }{C_{bat}}
\end{equation}
\parindent0pt
\begin{equation}  SoC_{avg,i,v} = SoC_{i-1,v} + \frac{ 0.5 \cdot I_{i,v}  \ \cdot \Delta t }{C_{bat}} \end{equation}
\parindent0pt
\begin{equation} SoC_{dev,i,v} =  \frac{ 0.5 \cdot I_{i,v} \cdot \Delta t }{C_{bat}} \end{equation}
\parindent0pt
\begin{equation} Ah_{i,v} =  \ I_{i,v} \cdot \Delta t \  \end{equation}

The weight $\alpha_3$ multiplies the ride availability objective, which is the negative product of decaying weights, $w_{i,v}$  multiplied to the corresponding power allocated. When rides are abound and EVs need to be charged quickly to service them in the order of earliest to latest deadlines the power allocated to the different vehicles are weighted using monotonically decreasing weights, $ w_{i,v} $ as given in equality constraint (18).

\begin{equation}
 w_{i,v} = \frac{1}{ i + TT_{v}} \ \ \ \forall v, \ i \in [0,TT_{v} - 1] 
\end{equation}

Minimizing the total negative power allocated to all the vehicles in all their time slots is equivalent to minimizing the total charging time. The decreasing weights are distributed throughout the time slots for each vehicle. The size of the decreasing weights are influenced by the charging period $TT_v$. The charging period is divided into uniform time slots, $i$ for each vehicle, $v$ and the cost of each of the three objective is computed for each vehicle at each time slot. The total cost for each objective is summed over all the time slots and vehicles in the schedule. These individual objectives are normalized using utopia and nadir points and then summed together as weighted sum of objectives as given in (8). The solution of the optimization results in the current allocation, $I_{i,v}$ in each time slot for each vehicle. The power allocated in each time slot is simply the product of the current allocation, $I_{i,v}$ to the constant charging voltage, $V$.

\section{Simulation and testing}

\subsection{Case study}

The Adaptive Charging Network, ACN \cite{b25} dataset contains real-world EV charging patterns at CSs across California Institute of Technology. The task information obtained from this data were connection time, disconnection time, energy requested in KWh, minutes available for charging and charging slot ID. Low and high vehicle charging density scenarios from two different weeks were used to compare the baseline solution with our proposed multi-objective optimization solution. The low density charging scenario was between 2021-05-22 to 2021-05-28 and the high density charging scenario was between 2021-05-01 to 2021-05-07. The maximum number of vehicles to optimize at one time for the low density scenario was 4 and for high density scenario was 14. The electricity price during the two weeks was obtained from \cite{b23}. The simulation code can be found here \footnote{https://github.com/JosePeeterson/ACN\_caltech}.

\subsection{ Simulation Flow }
The simulation starts on the first day with the first arrival event of an EV. The task information is recorded and a schedule is generated with power allocated over the different time slots of the EV's charging period using either the baseline solution or the proposed solution. The schedule is then implemented and the states of the EV such as the remaining charging time and current SoC level are updated until another event such as an arrival or departure of an EV occurs. When this happens, the new task information is recorded together with the updated states of the existing EVs and a new schedule is generated. This process repeats until the end of the week. 

\subsubsection{ Scheduling strategy using Baseline Solution}

The baseline solution first computes the minimum of either the number of time slots required to fully charge the EV to a SoC level of 100\% or the number of time slots available till the EV's departure time for each EV. This policy then applies the maximum power at every time slot in the charging period of each EV.

\subsubsection{Scheduling strategy using Proposed Solution}

The proposed solution on the other hand has a global task admission system that is aware of the existing schedules at all CSs and it ensures that adding a new vehicle to a local CS results in a feasible schedule for all vehicles in that station. Otherwise, the vehicle is denied immediate charging or allocated to another CS.
The number of time slots remaining for each EV till departure time is computed. Together with the new task information and updated states of all the EVs the three objectives are individually optimized in order to obtain the respective utopia and nadir points. The respective utopia and nadir points are used in the normalization of the weighted sum of the three objectives. Gurobi solver is then used to perform optimization to generate the power allocations at each slot for all the EVs.

\subsection{ Analysis of results }

The Performance metrics given in table 2 are used to evaluate and compare the results from the proposed solution with the baseline solution for all the vehicles in the two weeks of low and high vehicle charging density.

\begin{table}[htbp]
\caption{Performance metrics}
\begin{tabular}{|ccccc|}
\hline

\multicolumn{1}{|c|}{\multirow{3}{*}{\textbf{\begin{tabular}[c]{@{}c@{}}Performance Metrics \\ (unity weights) \end{tabular}}}} 
& \multicolumn{2}{c|}{\textbf{\begin{tabular}[c]{@{}c@{}}Low Vehicle \\ Charging Density\end{tabular}}} 
& \multicolumn{2}{c|}{\textbf{\begin{tabular}[c]{@{}c@{}}High Vehicle \\ Charging Density\end{tabular}}} \\ \cline{2-5}
\multicolumn{1}{|c|}{} & \multicolumn{1}{c|}{\textbf{Baseline}} & \multicolumn{1}{c|}{\textbf{Proposed}} & \multicolumn{1}{c|}{\textbf{Baseline}} & \textbf{Proposed} \\ \hline
\multicolumn{1}{|c|}{\textbf{\begin{tabular}[c]{@{}c@{}}Total Charging\\ Cost (\$)\end{tabular}}} & \multicolumn{1}{c|}{120.0} & \multicolumn{1}{c|}{56.2} & \multicolumn{1}{c|}{924.9} & 274.7 \\ \hline
\multicolumn{1}{|c|}{\textbf{\begin{tabular}[c]{@{}c@{}}Total Battery \\ Capacity\\ Fading (Ah)\end{tabular}}} & \multicolumn{1}{c|}{0.098} & \multicolumn{1}{c|}{0.082} & \multicolumn{1}{c|}{0.95} & 0.54 \\ \hline
\multicolumn{1}{|c|}{\textbf{\begin{tabular}[c]{@{}c@{}}Total Amortized \\ Battery\\ Value Loss (\$)\end{tabular}}} & \multicolumn{1}{c|}{5.42} & \multicolumn{1}{c|}{4.54} & \multicolumn{1}{c|}{52.58} & 29.9 \\ \hline
\multicolumn{1}{|c|}{\textbf{\begin{tabular}[c]{@{}c@{}}Total Peak\\ Power Period (Hr)\end{tabular}}} & \multicolumn{1}{c|}{45} & \multicolumn{1}{c|}{30} & \multicolumn{1}{c|}{347.5} & 159.25 \\ \hline
\multicolumn{1}{|c|}{\textbf{\begin{tabular}[c]{@{}c@{}}Total Charging\\ Time (Hr)\end{tabular}}} & \multicolumn{1}{c|}{22.2} & \multicolumn{1}{c|}{25.4} & \multicolumn{1}{c|}{74.6} & 103.4 \\ \hline
\multicolumn{1}{|c|}{\textbf{\begin{tabular}[c]{@{}c@{}}Maximum\\ Optimization\\ Time (mS)\end{tabular}}} & \multicolumn{1}{c|}{0} & \multicolumn{1}{c|}{563} & \multicolumn{1}{c|}{0} & 772 \\ \hline
\multicolumn{2}{|c|}{} & \multicolumn{3}{c|}{\textbf{\begin{tabular}[c]{@{}c@{}}Proposed Solution\\ (High Vehicle Charging Density)\end{tabular}}} \\ \hline

\multicolumn{2}{|c|}{\textbf{\begin{tabular}[c]{@{}c@{}}Objectives with different \\ weights ($\alpha_1$, $\alpha_2$, $\alpha_3$)\end{tabular}}} 
& \multicolumn{1}{c|}{\begin{tabular}[c]{@{}c@{}} \textbf{(0.6,} \\ \vspace{0.5ex} \textbf{0.3, 0.1)} \end{tabular}} 
& \multicolumn{1}{c|}{\begin{tabular}[c]{@{}c@{}} \textbf{(0.3,} \\ \vspace{0.5ex} \textbf{0.6, 0.1)} \end{tabular}} 
& \multicolumn{1}{c|}{\begin{tabular}[c]{@{}c@{}} \textbf{(0.1,} \\ \vspace{0.5ex} \textbf{0.3, 0.6)} \end{tabular}} \\ \hline
\multicolumn{2}{|c|}{\textbf{\begin{tabular}[c]{@{}c@{}}Total Charging\\ Cost (\$)\end{tabular}}} 
& \multicolumn{1}{c|}{274.6} & \multicolumn{1}{c|}{377.8} & 1071.1 \\ \hline
\multicolumn{2}{|c|}{\textbf{\begin{tabular}[c]{@{}c@{}}Total Amortized \\ Battery\\ Value Loss (\$)\end{tabular}}} 
& \multicolumn{1}{c|}{30.0} & \multicolumn{1}{c|}{27.7} & 38.5 \\ \hline
\multicolumn{2}{|c|}{\textbf{\begin{tabular}[c]{@{}c@{}}Total Charging\\ Time (Hr)\end{tabular}}} 
& \multicolumn{1}{c|}{103.4} & \multicolumn{1}{c|}{109.3} & 74.6 \\ \hline
\end{tabular}
\end{table}

Each of the performance metrics is calculated from all EVs in the entire week with some vehicles charging simultaneously. Unity weights represent equal weights for all 3 objectives, i.e. $\alpha_1, \alpha_2, \alpha_3 = 1$. Overall, the proposed solution outperforms the baseline solution on all metrics except total charging time and maximum optimization time. The increase in total charging time is acceptable for the proposed solution as the schedule generated was feasible. The optimization times are less than 1 second so real-time performance is still guaranteed. The total charging cost and total battery capacity fading cost is lower for the proposed solution in both cases of low and high density. In the low density case, the proposed solution has 53.2\% lower charging cost and 16\% lower capacity fade for the proposed solution.  Total Amortized Battery Value Loss is computed for a 86 KWh battery that has a cost of \$135/KWh. The cost of the battery multiplied to the ratio of total capacity faded to the nominal capacity gives the value lost in dollars. The impact of the baseline charging strategy on the grid during the high density case is measured by the total peak power period metric which is the total time of peak power usage for all vehicles. Peak power period is defined as a time slot allocated with greater than 75\% of maximum power. For the low density case, this leads to a 33\% reduction in total peak power period. Fig. 2 shows the total peak power period for all the charging EVs at each optimization time for the high charging density week. The proposed solution shown in blue uses unity weights and the baseline solution is shown in red. The average of the total peak power period over all optimization time for the proposed solution is 1.8 hours. This is lower compared to 4.3 hours for the baseline solution.

\begin{figure}[htbp]
\centerline{\includegraphics[width=8cm,height=5cm]{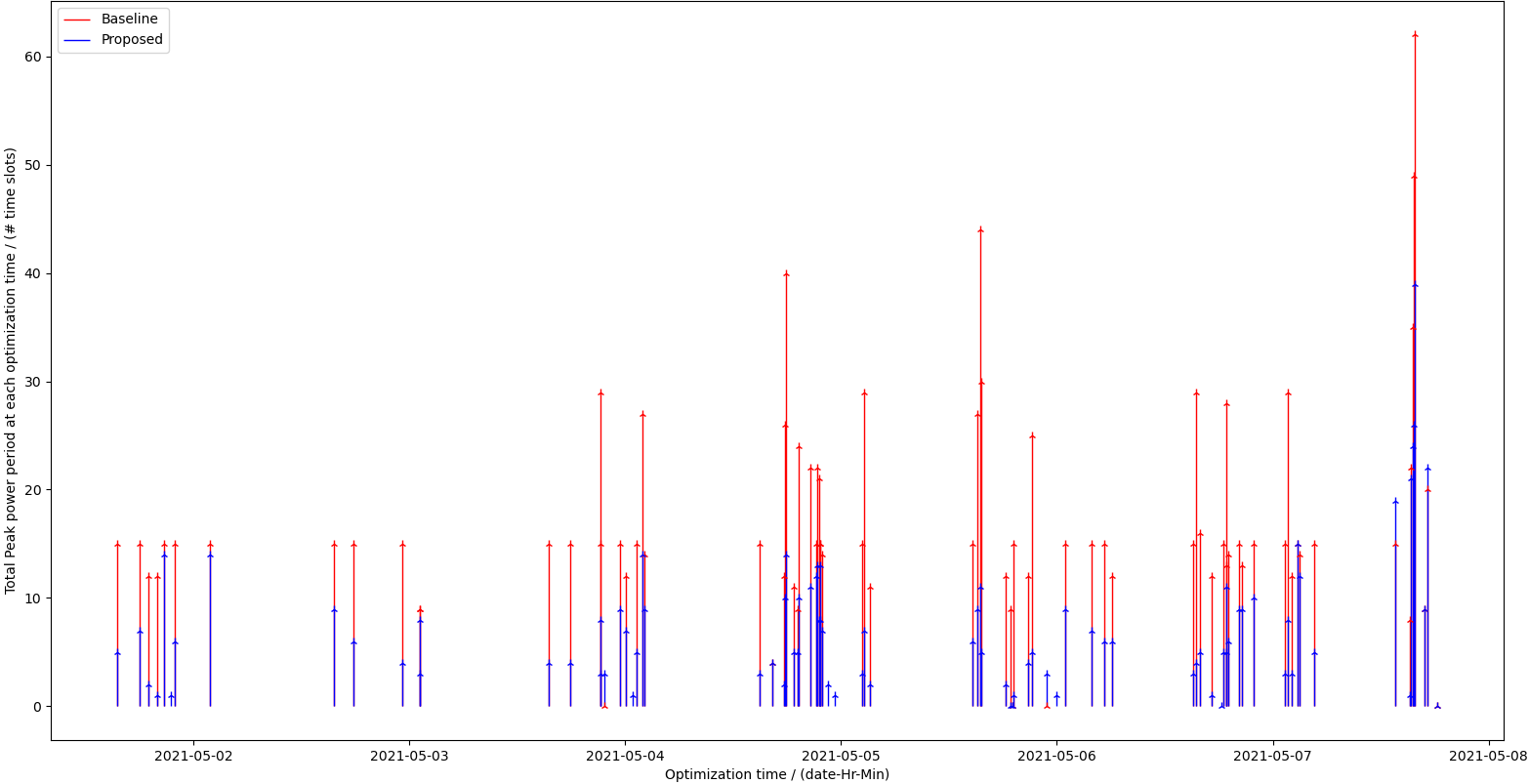}}
\caption{ Total Peak power period for baseline and proposed solution }
\label{fig}
\end{figure}

Fig. 3 shows one instance of the optimization using baseline and proposed solution taking place on 2021-05-07 at 15:44 for an arrival event of an EV. The top subplot shows the electricity price followed by the subplots for the charging current allocated to each vehicle at each time slot by the baseline solution and the proposed solution using unity weights. Each vehicle is given a distinct colour and their charging current values are represented as a stacked bar plot. For the baseline solution the schedule lasts only till 19:14 compared to the proposed solution that lasts till 01:59 on the next day. The maximum number of vehicles in this optimization are only 5 for baseline compared to 13 for the proposed solution. These observations are due to the maximum power allocation in successive optimization events for baseline solution whereas for the proposed solution charging is allocated closer to the departure and this results in aggregation for charging tasks that spill over into future optimization events. The baseline solution comes at the cost of charging during higher electricity prices, causing higher capacity fade and higher number of peak power periods. In contrast, the proposed solution has a monotonically decaying trend which can be attributed to the urgency created by the ride availability objective. Current peaking between 17:14 to 17:44 is the result of a drop in electricity price. The distribution of currents throughout the time slots till the end of the charging period of each vehicle and the two small current allocations near the end of the schedule for vehicle 0 are due to the battery capacity fade objective that tries to maintain lower SoC levels till the end of the charging period of each vehicle in the schedule.

\begin{figure}[htbp]
\centerline{\includegraphics[width=8cm,height=8cm]{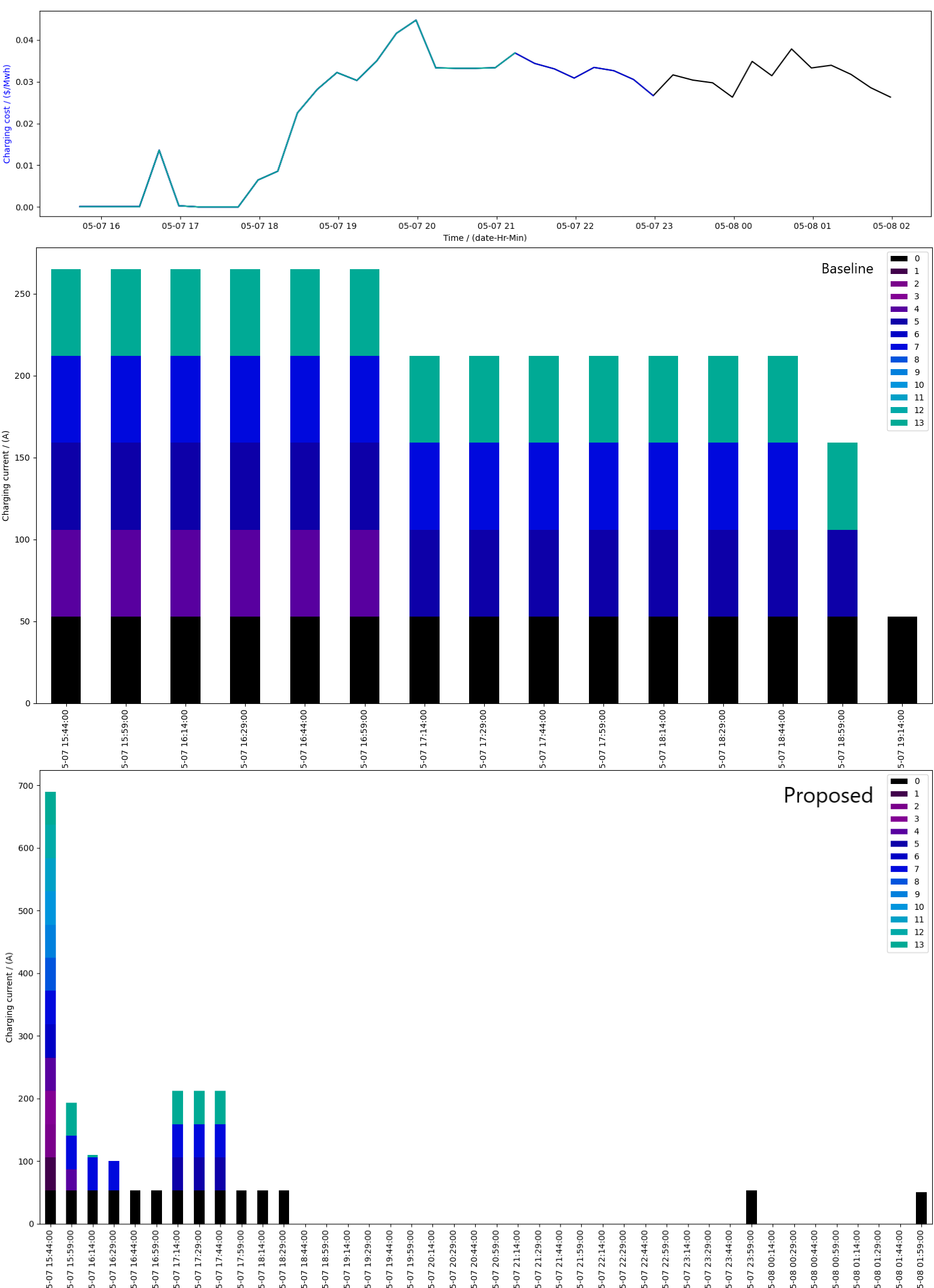}}
\caption{Stackplot of peak power time slots with baseline solution }
\label{fig}
\end{figure}

Although setting unity weights in conflicting objectives for every optimization event in the entire week leads to the deterioration of some objectives in some optimization events, overall the combined objectives is still better than the BAU policy of charging at the maximum rate as shown by the results in table 2. The lower part of table 2 also shows the effect of prioritizing one objective over another by setting higher weights. Clearly, setting a higher weight for the charging cost (0.6,0.3,0.1) has the most significant reduction in the charging cost compared to the other two weight combinations. Similarly, setting a higher weight for ride availability reduces the total charging time. The cost savings on battery value loss is not as significant for a single week of operation but in the long run it helps to mitigate the range anxiety problem and improve the resale value of the EV.\\

\section{Conclusion}
EVs are set to become an ubiquitous mode of transport by 2030. Many commercial EV fleet operators currently practice the BAU policy of charging their EV fleets at the maximum power at all times. However, there are many benefits in terms of monetary cost and electricity peak loading period reduction when the EV charging process is carried out in a controlled and coordinated manner. The EV charging cost was minimized through the forecasted electricity price. A detailed battery capacity fading cost was minimized for reasons of safety and range anxiety. The ride availability was maximized by minimizing the vehicle’s time spent on charging. Finally, The three objectives were formulated as a multi-objective optimization problem and solved for two scenarios from the real-world ACN dataset. The performance metrics indicate that the proposed choice of multi-objectives will benefit the fleet operator in saving  52.4\% more money than when using the baseline solution. The peak power period of the grid is also minimized by 33\%. This concludes that the proposed solution results in an improved charging policy compared to the BAU policy of the baseline solution.

\addtolength{\textheight}{-12cm}   




\end{document}